# Localized Nitrogen-Vacancy centers generated by low-repetition rate fs-laser pulses

Charlie Oncebay[1,2,‖], Juliana M. P. Almeida[1,3,‖], Gustavo F. B. Almeida[1,4],
Sérgio R. Muniz[1*], Cleber R. Mendonça[1*]

[1] *São Carlos Institute of Physics, University of São Paulo, 13560-970, São Carlos, SP, Brazil*
[2] *National University of Engineering, Science Departament, Peru.*
[3] *Department of Materials Engineering - Federal University of São Carlos, 13565-905, São Carlos-SP, Brazil*
[4] *Federal University of Uberlândia, Institute of Physics*

[‖] C.O. and J.M.P.A. contributed equally to this work.

*corresponding authors: crmendon@ifsc.usp.br; srmuniz@ifsc.usp.br

## Abstract

Among hundreds of impurities and defects in diamond, the nitrogen-vacancy (NV) center is one of the most interesting to be used as a platform for quantum technologies and nanosensing. Traditionally, synthetic diamond is irradiated with high-energy electrons or nitrogen ions to generate these color-centers. For precise positioning of the NV centers, fs-laser irradiation has been proposed as an alternative approach to produce spatially localized NV centers in diamond. However, most of the studies reported so far used high-repetition rate fs-laser systems. Here, we studied the influence of the irradiation conditions on the generation of $NV^-$. Specifically, we varied pulse fluence, laser focusing, and the number of pulses upon irradiation with 150 fs pulses at 775 nm from a Ti:sapphire laser amplifier operating at 1 kHz repetition rate. Optically Detected Magnetic Resonance (ODMR) was used to investigate the produced NV centers, revealing a sizeable zero-field splitting in the spectra and indicating the conditions in which the lattice strain produced in the ablation process may be deleterious for quantum information applications.

## 1. Introduction

There is much interest in developing methods allowing the precise placement of engineered solid-state quantum systems, such as manufactured artificial atoms (like spin color-centers and quantum dots), in different materials, for quantum technology applications. One of the most promising solid-state optically active spin systems currently studied is the nitrogen-vacancy (NV) center in diamond, which is usually produced by irradiating ultrapure synthetic diamond with beams of electrons or nitrogen ions. Recently, optical microfabrication using fs-lasers has been studied as an alternative method to create such defects, but most studies investigated systems with high-repetition rates, and few do a detailed study of the irritation conditions.

The NV center is one of the several hundred naturally occurring defects in the diamond lattice [1-3]. It is a point-defect created by a missing carbon next to a substitutional nitrogen impurity in the lattice, and it has attracted a lot of interest recently because of its distinct physical properties and many potential applications. For example, NV centers can be optically initialized in a well-defined quantum state, presenting long coherence times at room temperature (ranging up to a few ms in single NV) and allowing creating protocols to manipulate its spin-state using a combination of optical and magnetic resonance techniques. Finally, the system's quantum state can easily be read out by its [4-7], allowing to control and detect a single NV center with high efficiency. Such aspects have opened the possibility of exploring ideas and applications of quantum information and quantum computing at room temperature [7-11].

NV centers can also be used as quantum sensors to detect physical parameters, such as temperature, magnetic and electric fields [12-15], which have been explored even in studies of biomolecules and cell structures [16]. Furthermore, NV centers in diamonds have emerged as a promising platform for integrated photonic circuits [17-19], combining their nonlinear optical properties with excellent thermal properties. The nonlinearity of diamond [20-22] has be demonstrated to be useful for frequency conversion in the wavelength range of telecommunication, for example [23].

In this context, the controlled production of single NV centers, or a controlled set (ensemble) of them [24, 25], is of utmost importance to developing semiconductor-based quantum devices because the ability to engineer such devices can be widely used in quantum optics, magnetometry, thermometry, and other applications. Therefore, such controlled production has prompted the development of new methods that can be potentially used for the precise positioning of an individual NV center, or a small set of them, in particular regions of a diamond structure.

To generate the color-centers, the synthetic diamonds are typically irradiated with high-energy particles, from a few keV to MeV, usually with electrons or nitrogen ions. In addition to the particle energy, another tunable parameter is the irradiation dose, which determines the density of vacancies generated in the sample [26-28]. Alternatively, ultrashort-pulse laser irradiation has also been proposed as a potential method for generating color centers, vacancies and NV centers in diamonds [29-33]. In this approach, intense laser pulses propagating in air ionizes molecules, like $O_2$ and $N_2$, generating free electrons that are accelerated by the subsequent pulses to produce a beam of electrons that collide with carbon atoms at the diamond's surface, taking them out of the crystal lattice, thus generating the vacancies. These vacancies can move during annealing (at high temperatures), creating the possibility of binding to some nitrogen impurity in the crystal, resulting in the NV center [29]. Direct generation of vacancies by fs-laser followed by annealing to promote $NV^-$ centers have also been reported [31].

Preliminary investigations on the subject demonstrated diamond transformation to amorphous carbon, graphitic phases and 3H center formation when using picosecond-laser irradiation [33]. Later, the same team reported the enhancement of NV luminescence in diamond due to $sp^3$ carbon allotropic structures induced by picosecond laser exposure [30]. When using fs-laser, it was demonstrated the formation NV centers in high nitrogen content-diamond through photoluminescence (PL) measurements [29]. More recently, similar achievements was reported in diamond with nitrogen concentration < 5ppb by using a statistical approach and single fs-laser pulses [31] or using UV fs-laser pulses at nanoablation regime, in which laser fluence was kept below a certain threshold [32].

Although ultrashort-laser has been demonstrated to be useful for NV center generation in diamonds, further investigations are essential to determine the formation of such centers by femtosecond pulses at low-repetition, particularly using more precise techniques, including those that enable magnetic sensing by electronic spin resonance.

Here, we demonstrate through optically detected magnetic resonance (ODMR) the generation of active and spatially localized NV centers in diamond by using fs-laser pulses at low repetition rate, while causing the least damage to the sample. Additionally, we determined the damage threshold fluence for the diamond surface, the influence of laser beam focusing, and the number of pulses on the center generation. The generation of NV centers was confirmed by Raman spectroscopy, photoluminescence, and ODMR. Our results show that the ODMR spectrum has a broad dual-dip separated by ~25 MHz without an applied external magnetic field, possibly related to the lattice strain caused by fs-laser irradiation, which may influence the use of such defects for quantum information.

## 2. Material and Methods

Experiments were performed on a 5×5×1 mm$^3$, type-Ib synthetic diamond, produced via the chemical vapor deposition (CVD) method, with nitrogen impurity level below 1 ppm. To generate defects in the diamond sample, we used 150 fs pulses from a Ti:sapphire (Clark-MXR) laser at 775 nm, with a repetition rate of 1 kHz. An objective lens focused the laser beam onto the sample, mounted on a three-axis computer-controlled motorized positioning stage used to translate the sample at a constant speed (10 μm/s). After fs-laser irradiation, the sample was annealed at a temperature of 680 °C for 30 minutes to remove the amorphous carbon generated in the ablation process and to increase the mobility of the vacancies possibly generated during the irradiation. Finally, the sample was cleaned with a mixture of acids (hydrochloric acid, nitric acid, and sulfuric acid in the proportion of 1: 1: 1) to remove any remaining impurities from the surfaces.

The samples were analyzed in a commercial confocal microscope system (Zeiss LSM-780) under laser excitation at 543 nm. In addition, Raman measurements at room temperature were carried out with a Witec micro-Raman (model Alpha 300S A/R), using an argon laser (Melles Griot, model 35-LAL) at 514 nm and 100x objective lens for excitation and grating of 600 grooves.mm$^{-1}$. The integration time was set at 1 s and the laser power at ~8 mW.

The fluorescence spectroscopy of the NV center was done in a homebuilt setup that uses a laser operating at 532 nm as the excitation source. First, the 532 nm beam was expanded and collimated in a telescope before being focused by a microscope objective (100x/1.4NA or 50x/0.95NA). Next, the emitted photoluminescence by the NV centers is collected by the same microscope objective, filtered with a 561 nm dichroic mirror and a long-pass filter with a cutoff at 550 nm, to remove the 532 nm excitation light. This combination allows blocking all residuals of green laser light while keeping most of the light emitted by the NV center. Finally, the collected and filtered photoluminescence is measured in either a CCD camera, a photomultiplier, or a spectrometer, depending on the experiment's purpose. In this setup, the position of the sample is controlled with a sample holder mounted on a nanopositioning stage (Thorlabs), which enables the translation in three orthogonal axes.

Additionally, we conducted a microwave spectroscopy study of the produced NV centers using optically detected magnetic resonance (ODMR) in the same homemade optical setup. This method detects the electron spin resonance (ESR) in the $NV^-$ center by examining the change in fluorescence captured by the microscope while sweeping the microwave field across a frequency range around 2.8 GHz. More specifically, the frequency of the microwave field (MW) is swiped around the ground state energy of the NV center ($^3A$) while the fluorescence resulting from illumination by a laser exciting the $^3A \rightarrow ^3E$ transition is monitored [34, 35]. For the microwave, we used a setup consisting of a signal generator (Standford Research, SG384) connected to a computer-controlled fast switch to control the MW pulse used for the excitation of the NV electron. All the ODMR spectra were obtained without any externally applied magnetic field and with the sample maintained at a temperature of around 300 K.

### 3. Results and Discussion

Initially, fs-laser pulses were used to produce lines in the diamond sample, varying the pulse energies ($E$) to determine the damage threshold energy ($E_{th}$). To focus the fs-laser, we employed a 1.25 NA objective. During the process, the sample was moved at a constant speed of 10 µm/s. Figure 1 shows the squared values of the measured half-width ($r$) for the fabricated lines as a function of the pulse energy (log-scale). The values of $r$ were determined using optical microscopy images. According to the zero-damage method [36], by fitting the data with $r^2 = \frac{w_0^2}{2} ln\left(\frac{F}{F_{th}}\right)$, in which $w_0$ is the beam radius at the focus, one obtains $F_{th}$. The line in Fig. 1 represents the best fitting curve, from which we determined the threshold fluence $F_{th} = 1.3 \pm 0.1 \text{ mJ/cm}^2$

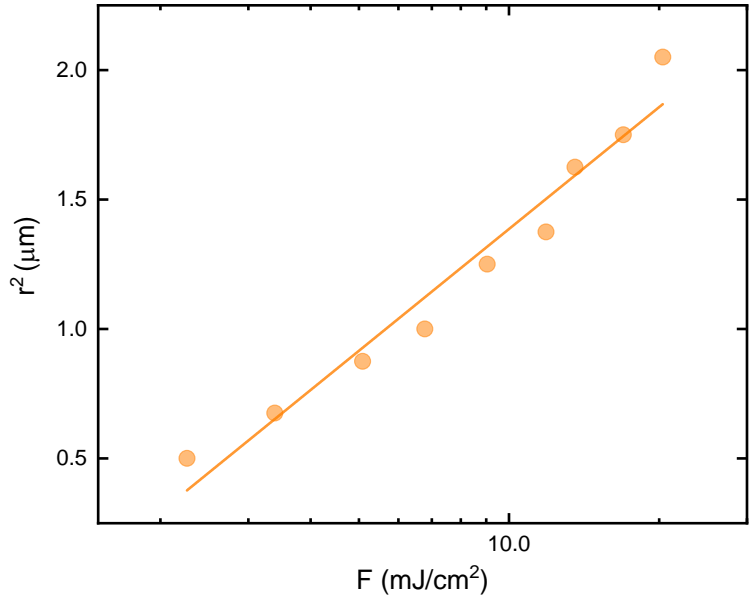

**Figure 1:** Squared values of the measured half-width of micromachined lines as a function of the pulse fluence (log-scale) and its best-fit (line), according to the expression given in the text.

Under the previously described conditions, the lines produced on the diamond were subsequently analyzed in a fluorescence confocal microscope (Zeiss LSM-780), using a laser operating at 543 nm as the excitation source. As shown in Fig. 2, the lines display the fluorescent emission in several regions due to optically active defects generated upon fs-laser irradiation. The fluorescence was observed only for lines irradiated with pulse fluence higher than approximately 11 mJ/cm². However, for pulse fluence higher than 34 mJ/cm² the damage on the diamond's surface started to get too severe, also resulting in less fluorescence. Thus, such results indicate the pulse fluence range 11 – 34 mJ/cm² as the best option to generate active defects with the least damage to the surface. Fig. 2 shows some examples of the density of defects generated as the pulse energy increases.

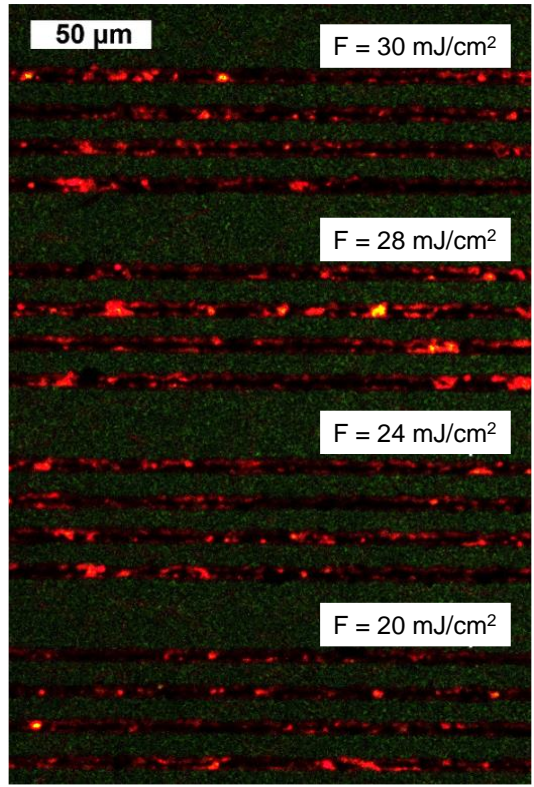

**Figure 2:** Confocal microscopy image (excitation at 543 nm) of lines irradiated with different fs-pulse energies. Fluorescence (600 – 700 nm) spots are due to defects created by the fs-laser and seem to mostly increase with the energy.

Aiming to generate defects with minimal damage to the surface, we did the fs-laser irradiation with a fixed pulse fluence of 14 mJ/cm$^2$ and varied the microscope's objective position with respect to the surface. For that, we started with the focal point at 50 μm below the surface and moved it to a distance of 20 μm above the surface, in steps of 10 μm for each line. In this study, we used an objective with NA=0.85 and maintained the lateral speed of the sample at 10 μm/s. Figure 3 shows confocal microscopy images for some relevant focal distances. One can observe that when the laser focus is positioned 20 μm above the surface (Z = +20 μm), the damage is significantly lower than in the other lines, as shown in the microscopy image of the diamond surface in Fig. 3A. Still, there are NV centers created under this condition, as shown in the fluorescence of Fig. 3B. At Z = +10 μm, a lot of fluorescent defects can still be observed. For the conditions described here, our results indicate that an optimal irradiation condition for obtaining active defect centers with minimal surface damage corresponds to a pulse energy of 14 mJ/cm$^2$, with the focal point located between 15 μm and 10 μm above the sample surface.

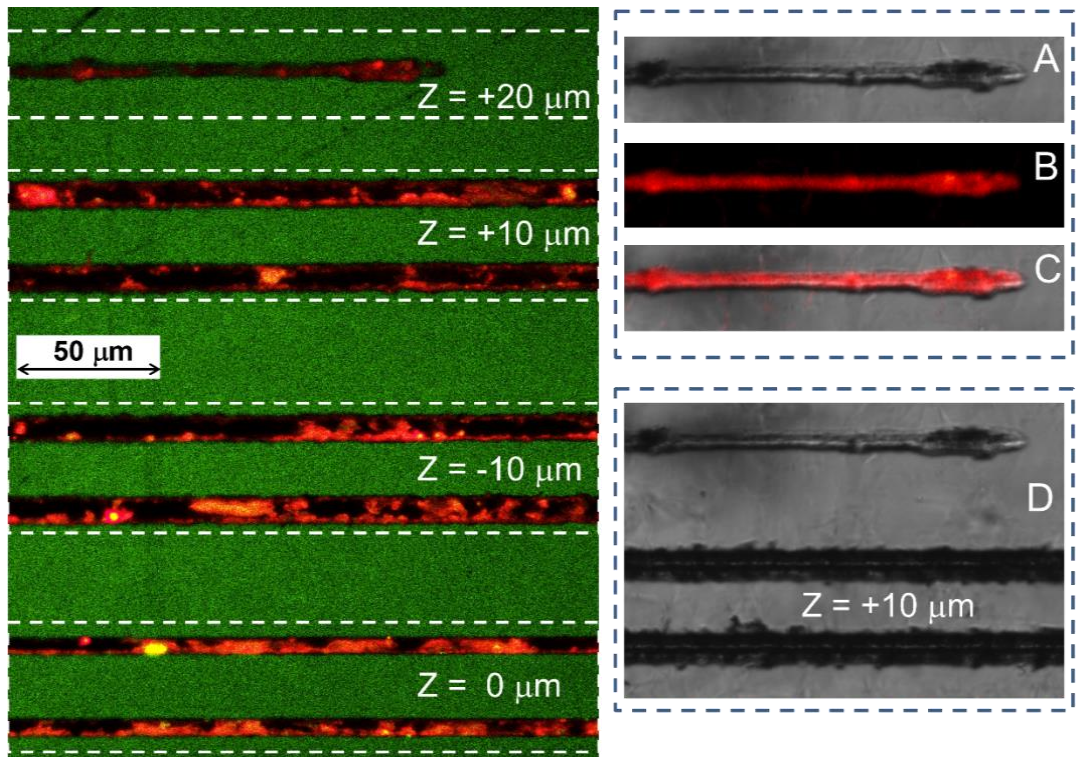

**Figure 3:** Confocal microscopy images of lines made with pulse fluence of 14 mJ/cm$^2$ and laser focused on different positions. The excitation wavelength is 543 nm. Positive Z indicates that the laser focus is above the sample surface, while negative means focus below the surface. In the box on the right, the figures show the region of Z = 20 μm observed with: (A) regular white-light microscopy, and (B) only fluorescence microscopy ($\lambda_{fluo} > 600$ nm; $\lambda_{excit} = 543$ nm), while (C) shows a composition combining images (A) and (B). For comparison, the box on the bottom-right shows part of the bright-field image to compare the damage produced on the surface at Z = 10 μm and 20 μm. Images (A)-(D) were taken before the final cleaning stage with the acids, described in section 2, while the image on the left was after the acid cleaning.

In addition to the fluorescence microscopy, we used micro-Raman spectroscopy to verify the presence of NV centers in the lines fabricated by fs-laser micromachining, as displayed in the solid-line spectrum in Fig. 4. The dotted line in Fig. 4 corresponds to the sample's pristine (non-irradiated) reference area. The spectrum of the pristine diamond features a well-known sharp peak at 1332 cm$^{-1}$ due to the sp$^3$ carbon bonds, also called the first-order Raman peak of diamond [32, 37-39]. No significant change was observed in this peak after fs-laser irradiation. However, minor changes were detected between 2000 and 3500 cm$^{-1}$, as displayed in the inset of Fig. 4.

The small peak at 2010 cm$^{-1}$ (corresponding to 574 nm) is attributed to the zero-phonon line (ZPL) of the NV center in the neutral charge state (NV$^0$) [39, 40]. The enhancement of this peak has been reported with the increase of nitrogen impurity content in CVD diamond for the amounts of 20 and 40 ppm [39]. Although minor peaks related to defect centers have been observed in the Raman spectra of diamond after fs-laser micromachining, the negatively charged NV center (NV$^-$) was not detected (one would expect a Raman peak at around 3750 cm$^{-1}$, i.e., 637 nm [40]). This might be related to the low concentration of NV$^-$ or because the excitation source used in Raman measurements (514 nm) is not resonant with energy levels associated with NV$^-$ centers. [40].

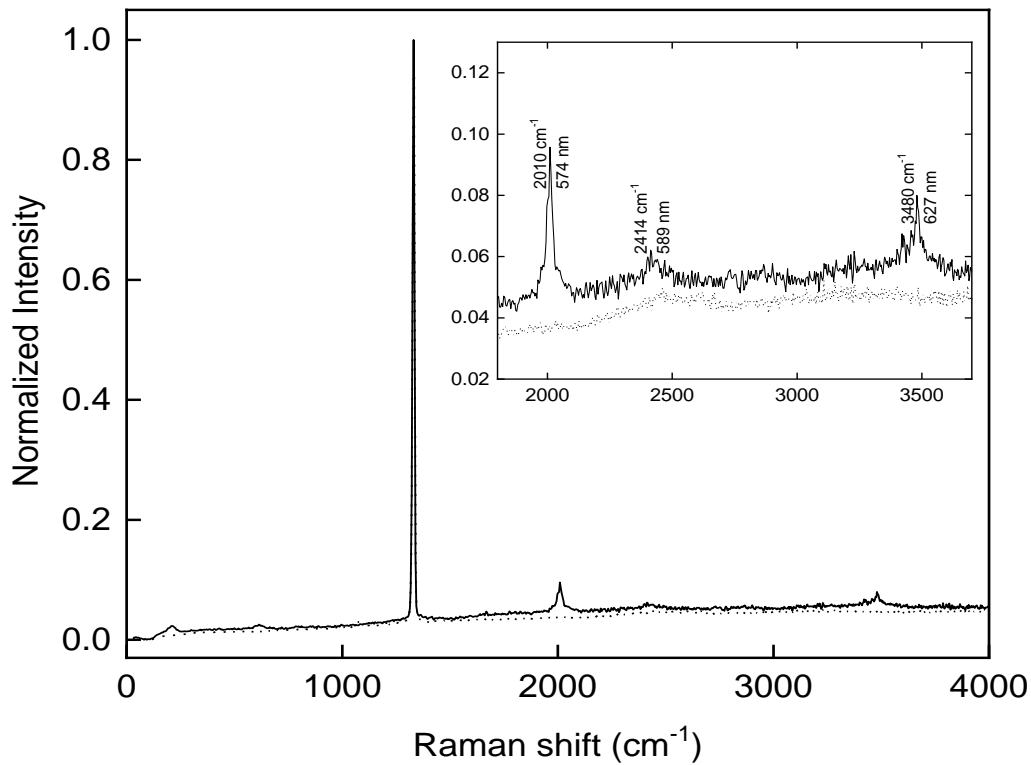

**Figure 4:** Raman spectroscopy (514 nm) of the pristine sample (dotted line) and fs-laser (solid line). The inset shows a detail of the region between 2000 and 3500 cm$^{-1}$.

Based on the previous results, we studied the production of NV centers as a function of the number of fs-pulses. For that, we irradiated a fixed spot in the sample using the same microscope objective (NA=0.85), with the focus positioned at $Z = 15 \pm 1$ μm above the sample surface and kept the pulse fluence at 14 mJ/cm². Figure 5 shows the results for 2k, 4k, 8k, 16k, and 32k pulses. Fig. 5(a) shows a photoluminescence image of the irradiated spots obtained with the homebuilt microscope under 532-nm laser excitation. The signal corresponds to the spectral integration of the NV center emission (600 – 800 nm). As it can be seen, luminescence is observed for 2k, 4k and 16k pulses, the latter being the most intense one.

On the other hand, no emission has been observed for 8k and 64k pulses, and a subtle signal was observed for 32k. These results show that the density of defects grows with the number of pulses, but for some areas of the sample, no NV centers are generated regardless of the irradiation conditions. This was observed before and is consistent with a stochastic process, in agreement with the literature [41]. It depends not only on the irradiation parameters but also on the local structure of the sample and/or the presence of seed nitrogen atoms. Figure 5(b) displays the emission spectra of the NV center (532 nm excitation and 300 K) for the regions irradiated with 2k, 4k, and 16k pulses, corresponding to Fig. 5(a), collected with an optical fiber and measured in a fiber-coupled spectrometer. Fig 5(b) shows the spectra of the broadband photoluminescence, where one can notice the characteristic 637 nm (1.945 eV) zero-phonon line of the $NV^-$ centers.

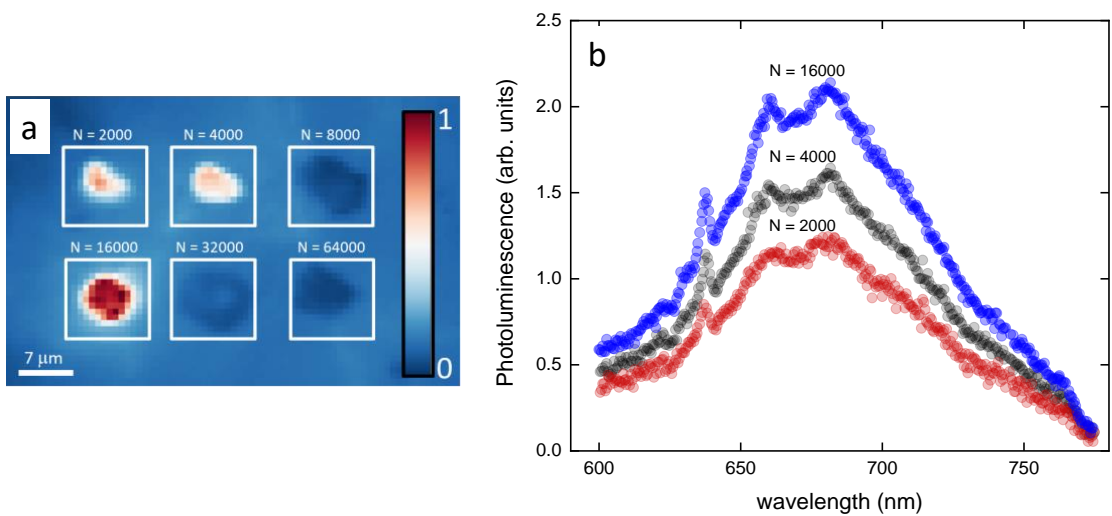

**Figure 5:** (a) Room-temperature photoluminescence images from an irradiated sample (scale bar 7 μm) at different numbers of pulses, and (b) corresponding photoluminescence spectra. Laser excitation at 532 nm.

As a final study, Fig. 6 shows the zero-field ODMR spectra (300K) obtained for the regions indicated in Fig. 5(a), irradiated with different numbers of pulses from the fs-laser. These spectra confirm the presence of the produced $NV^-$ centers. As expected, one sees that when the microwave frequency is on resonance, at around 2870 MHz, the ground state population of the NV center is transferred to $m_s = \pm 1$, resulting in a slight reduction of the fluorescence. However, the ODMR signal in Fig. 6 presents a quite broad dual-dip, even without an external magnetic field. We attribute this feature to a permanent lattice strain caused during the fs-laser irradiation [42, 43].

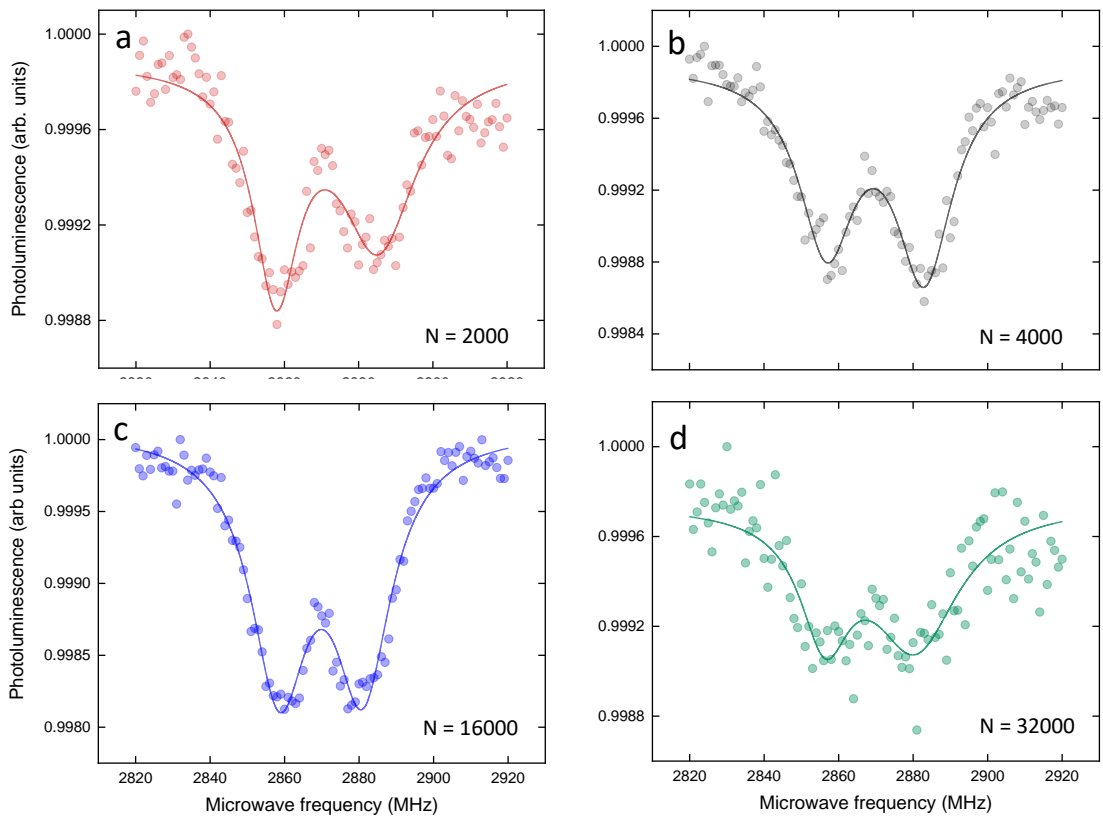

**Figure 6** – Optically detected magnetic resonance (ODMR) spectra measured at zero-field and room-temperature for the highlighted regions in Fig. 5(a).

Therefore, depending on the experimental conditions used in the production of localized $NV^-$ centers, considerable lattice stress can also be induced, affecting the application of such defects for quantum information.

## 4. Conclusion

The results reported here demonstrate the generation of $NV^-$ center when 150 fs pulses (775 nm) with fluences higher than 11 mJ/cm$^2$ are employed. For the parameter used in this study, we observed that the conditions for creating active NV centers with minimum surface damage were obtained for 14 mJ/cm$^2$ of pulse fluence and with the laser beam focus positioned 15 to 20 $\mu$m above the sample surface. Furthermore, the density of the NV center was shown to grow with the number of fs-pulses. However, the generation process is stochastic; for certain regions, no NV centers are generated independently of the irradiation conditions. Finally, from the zero-field ODMR, we observed that even minimizing the surface damage, the local strain caused by the laser irradiation may affect the electronic states of the NV center, shifting and broadening the electron-spin resonance. This effect may jeopardize its use in certain quantum information devices. Therefore, further studies are necessary to optimize the controlled production of active NV centers while minimizing the deleterious effects on the diamond crystal and the electron state properties.

## Acknowledgments

The authors acknowledge FAPESP (grants 2018/11283-7, 2019/27471-0, 2015/17058-7, 2013/07276-1, and 2009/54035-4), Air Force Office of Scientific Research, CNPq, and CAPES for financial support.